\documentstyle[prl,aps,preprint]{revtex}
\tightenlines

\newcommand{\be}{\begin{equation}}
\newcommand{\ee}{\end{equation}}
\newcommand{\bea}{\begin{eqnarray}}
\newcommand{\beas}{\begin{eqnarray*}}
\newcommand{\eea}{\end{eqnarray}}
\newcommand{\eeas}{\end{eqnarray*}} 
\newcommand{\ba}{\begin{array}}
\newcommand{\ea}{\end{array}}

\begin{document}

\draft
\preprint{\vbox{\hbox{UMD-PP-01-050}\hbox{IFP-795-UNC}\hbox{April 2001}}}

\bigskip
\bigskip

\title{ Prediction of $sin^2\theta_W$ in a Conformal Approach to
Coupling Unification} 
\author{P.H. Frampton$^1$\footnote{frampton@physics.unc.edu}
R. N. Mohapatra$^2$\footnote{rmohapat@physics.umd.edu},
Sungwook Suh$^1$\footnote{swsuh@physics.unc.edu}}
\address{$^1$ Department of Physics, University of North Carolina, Chapel
Hill, NC 27599\\
$^2$ Department of
Physics, University of Maryland, College Park, MD, 20742}


\maketitle
\begin{abstract}
{The possibility that non-supersymmetric conformal field theories softly
broken below  100 TeV may provide an alternative to conventional grand
unification is explored. We consider a low energy theory presumed to be of
this type arising from the
Type IIB superstring compactified on a $AdS_5 \times S_5/\Gamma$
space whose gauge group and the particle content are severely
restricted by the compactification process. We present an example of a
resulting $SU(4)_C\times SU(2)_L\times SU(2)_R$ with three generations,
which leads to coupling unification and a prediction for
$sin^2\theta_W\simeq .227$ and other phenomenology generally consistent
with observations.}
 \end{abstract} 

\bigskip
\bigskip


\section{Introduction}
The idea that all gauge couplings describing the forces in the standard
model may eventually merge into one is not only an aesthetically pleasing
one, but also has the virtue of making testable predictions at low
energies\cite{pati}. Two predictions typical of simple grandunification
models are $sin^2\theta_W (M_Z)$ and proton life time. The earliest grand
unified theories however had the problem of an enormous gauge hierarchy
that separated the weak scale from the scale of grand unification. The 
introduction of supersymmetry
not only provided a solution to this problem but also
provided a simpler framework for unification of couplings. While this is
perhaps the simplest approach to coupling unification, at present there is
no experimental evidence for this view. It is therefore interesting to
pursue other approaches to coupling unification and explore the
possibility of constructing phenomenologically viable models and isolate
their tests.

Motivated by conformal invariance of string theories, a new
approach to coupling unification has been suggested in a series of recent
papers\cite{paul1}. The basic idea of this approach relies on the
fact that a type IIB string theory compactified on an $AdS_5 \times S_5$ 
gives rise to an ${\cal N}=4$ SU(N) gauge theory, which has been known
for some time\cite{mandel} to be conformal due to extended global symmetry
and nonrenormalization theorems. All of the beta functions in this theory
vanish, making the couplings ``static''. The $\cal N$=4 supersymmetry is
crucial for the conformal invariance. However, it has been shown that by
compactifying the theory on $S_5/\Gamma$, where $\Gamma$ is a discrete
subgroup of the R-symmetry group SU(4), one can obtain conformal invariant
SU(N) gauge theories with finite N and lower number of
supersymmetries\cite{kachru,vafa}. One could then add soft conformal
breaking
terms in order to introduce the mass scales, say M. The couplings below
the scale M will then ``run'' following the usual rules of field theories.
If the constraints of $\sin^2\theta_W$ and compactification then allowed
for unification of the gauge couplings, then one could consider these
theories as alternatives to conventional grand unified theories. In
particular, if the scale M is in the TeV range, then one may completely
bypass low energy supersymmetry. In \cite{paul2}, it was shown how to use the abelian
$\Gamma = Z_7$ to arrive at $\sin^2 \theta_W(\mu) = 3/13 = 0.231$;
here we examine the situation for non-abelian $\Gamma$.

In a recent paper, an exhaustive classification of possible gauge groups
and the particle contents that arise from different compactifications of
AdS/$S_5$ have been presented\cite{keph}. In this note, we study one
particularly interesting class of models based on the gauge group
$SU(4)_C \times SU(2)_L\times SU(2)_R$ and show that this model can lead to 
prediction of $sin^2\theta_W$ at the $M_Z$ scale in agreement with
experiment. The nontriviality of this result follows from the fact that
group theory and unification allows only a small number of possibilities
for the value of $sin^2\theta_W$ at the conformal scale $M$, which we
choose in this case to be of order 100 TeV to be consistent with low energy rare
weak processes such as $K_L\rightarrow e^+\mu^-$.

\newpage

\section{The model}
We start with an AdS/$S_5$ compactification of a type II B string theory
which has ${\cal N}=4$ supersymmetry and is conformal invariant. This
theory has an SU(4) global
symmetry as its R-symmetry. We start with two D-branes in this theory
at a point in $S_5$ so that the starting gauge theory is $SU(2)$. We then
consider the nonabelian discrete group $\Gamma = Z_3\times D_4$ which is a
group of order 24. Recall that the group $D_4$ consists of the eight
rotations that leave a square invariant; two of the rotations are flips
about two lines that bisect the square and the other four are 90, 180, 270
and 360 degree rotations about the perpendicular to the plane of the
square. Following the conjecture by Lawrence et al\cite{vafa}, we assume
that the gauge group of this theory is $\Pi_i SU(2d_i)$, where $d_i$ are
representations of the $\Gamma\equiv Z_3\times D_4$. The representation of
$\Gamma$ are three 2-dim ones and 12 1-dim. which leads to the low energy
gauge group $[SU(4)]^3\times [SU(2)]^{12}$. A diagonal sum of this
group can be chosen as the actual gauge group of the model. In particular,
we will consider various diagonal sums of the type $SU(4)_C\times
SU(2)_L\times SU(2)_R$. Here SU(4) is the diagonal subgroup
of $r$ of the 
SU(4)'s where $r = 1$ or $2$ since, as shown in
\cite{keph}, $r=3$ disallows chiral fermions.
The two SU(2)'s are direct sums of $p$ and $q$ of the original
SU(2)'s such that $p+q=12$. The model is left-right symmetric\cite{lr} 
with quarks and leptons unified via $SU(4)_c$ a la Pati and Salam. It then
follows trivially that the resulting
gauge couplings of the effective SU(4) and the SU(2)'s respectively are:
$g_4=g_U/\sqrt{2 r}$ (where we insert a
factor $(1/\sqrt{2})$ for the dimensionality $d_i$ of the representation)
; $g_{2L}= g_U/\sqrt{p}$ and
$g_{2R}=g_U/\sqrt{q}$, with $p+q = 12$. We have assumed that the
gauge couplings become unified at the conformal
scale and due to conformal nature of the theory they remain frozen
afterwards. For the case of ${\cal N} =0$ supersymmetry, this is a conjecture, which
has been checked only to one loop. We assume this to be true to all
orders. Note that since $p$ and $q$ are integers, it is apriori not at
all obvious that this will lead to an acceptable prediction for
$\sin^2\theta_W$ at the weak scale. As we show below however, for a
unification scale of the order of 100 TeV, one does indeed get an
acceptable value.

The fermion and the Higgs content of the group is given by the quiver
diagram of the group\cite{keph}. They include Higgs as well as fermions in
all bi-fundamentals. The fact that the embedding of
representations of $\Gamma$ into the {\bf 4} of SU(4) are not
real guarantees guarantees the existence of chiral fermions\cite{paul1}.
It has been shown that in the $Z_3\times D_4$ embediing into SU(4) that we
are considering, there exists precisely three fermion generations in the
bifundamentals: (4,2,1), $(\bar{4},1,2)$; any fermion in
(1,2,2) representation can be given a bare mass term and be made to
decouple from the low energy part of the spectrum.

The Higgs bosons are in the representations (4,2,1), (4,1,2), (1,2,2); let
us denote these by $\chi_L$, $\chi_R$ and $\phi$ respectively.

The question we address in this paper is whether with the above particle
particle
content, one can have a viable theory. The first point we check is the
prediction for $\sin^2\theta_W$ at the scale $M_Z$. Before proceeding to
this discussion, we briefly review how the three generation model with the
gauge group $SU(4)_C \times SU(2)_L\times SU(2)_R$ emerges in this picture.

\newpage

\section{Complex Scalars and Chiral Fermions}

\bigskip

Some explanation of the matter content of complex scalars and
chiral fermions will be useful to the reader.

\bigskip

The gauge group arising from the orbifold $S^5/\Gamma$ and $N$ branes
is, as already mentioned, $G = \otimes_i SU(Nd_i)$ where the $d_i$ are
the dimensionalities of all the irreps of $\Gamma$.

\bigskip

The content of matter fields can be found from the chosen embedding
of $\Gamma$ in SU(4), which may be specified from the
decomposition of the {\bf 4} of SU(4). The {\bf 6}
of SU(4) is then the antisymmetric product 
$({\bf 4} \times {\bf 4})_{antisymm}$.
For any consistent embedding $\Gamma \subset SU(4)$,
the {\bf 6} is necessarily real. 

\bigskip

Given the content of the {\bf 6} in terms of irreps of $\Gamma$
the representations of the complex scalars under the gauge group $G$
are derived by producting the {\bf 6}
with all irreps of $\Gamma$, using the multiplication
table, as provided in {\it e.g.} the Appendix of \cite{keph}.

Each term appearing in the decomposition of
such products leads to the survival of a bi-fundamental
representation of the corresponding $SU(Nd_i) \times SU(Nd_j)$.
In the special case $i=j$, this is to
be interpreted as an adjoint plus a singlet of $SU(Nd_i)$.
The procedure can be conveniently
summarized and facilitated by use of a quiver
diagram\cite{douglas}.

\bigskip

For chiral fermions, the same procedure is followed
using the {\bf 4} instead of the {\bf 6}.
To achieve survival of chirality it is necessary that the {\bf 4}
not be real or pseudoreal, as proved in \cite{keph}.
The chiral fermions can be derived by a quiver diagram
in which at least some of the ``arrows" are oriented (chiral).
This is in contrast to
the scalar quiver diagram in which 
all the arrows are non-oriented (self-conjugate).

These consideraions have been applied in ref.\cite{keph1} to show that
the gauge group at the conformality scale is $[SU(4)_1\times SU(4)_2\times
SU(4)_3\times \otimes_i SU(2)_i$ with $i= 1,2, ... 12$. The fermion
content of this model consists of 

\noindent a.) Below we list the 24 bifundamentals $(2_i,2_j)$ under the
$i$th and $j$th SU(2)'s and singlets under the rest of the groups. (We
will employ this notation throughout this paper; below the bold face
numbers to the left of a bracket denote how many such multiplets are
there.);

\begin{eqnarray}
{\bf 2}(2_1,2_2), (2_1, 2_3),(2_1,2_5); {\bf 2}(2_3,2_4), (2_3, 2_5);
{\bf 2}(2_5, 2_6); (2_2,2_4), (2_2,2_6); (2_4,2_6);\nonumber\\
 {\bf 2}(2_7,2_8),(2_7, 2_9), (2_7, 2_{11}); {\bf 2}(2_9, 2_10), (2_9,
2_{11}), {\bf 2}(2_{11},2_{12}), (2_8,2_{10}), (2_8, 2_{12}), (2_{10},
2_{12}). \nonumber
\end{eqnarray}

\bigskip

\noindent b.) three ${\bf 2}(4_i,\bar{4}_i)$'s under the SU(4)'s and
$(4_1, \bar{4}_2), (4_2, \bar{4}_3), (4_3, \bar{4}_1)$ + c.c.

\bigskip

\noindent c.) 12 bifundamentals of the form $(2_i, 4_a)$ under
$SU(2)_i\times SU(4)_a$ given by: 
$(4_1, 2_5)\oplus (4_2,2_1) \oplus (4_3,2_3)\oplus (4_1, 2_6)\oplus
(4_2,2_2) \oplus (4_3,2_4)\oplus (4_1, 2_{11})\oplus (4_2,2_7) \oplus
(4_3,2_9)\oplus (4_1, 2_{12})\oplus (4_2,2_8) \oplus (4_3,2_{10})$

\bigskip

\noindent d.) Anti-bifundamentals 
$(\bar{4}_1, 2_4)\oplus (\bar{4}_2,2_6) \oplus (\bar{4}_3,2_2)
\oplus (\bar{4}_1, 2_9)\oplus (\bar{4}_2,2_11) \oplus (\bar{4}_3,2_7)
\oplus (\bar{4}_1, 2_{10})\oplus (\bar{4}_2,2_{12}) \oplus (\bar{4}_3,2_8)
\oplus (\bar{4}_1, 2_3)\oplus (\bar{4}_2,2_5) \oplus (\bar{4}_3,2_1)$.

\bigskip

Turning now to the scalar multiplets, we have the following:

1.) Bifundamentals of $(2_i, 2_j)$ type:

$(2_1, 2_4), (2_1, 2_6), (2_3, 2_2), (2_3, 2_6), (2_5, 2_2),(2_5,2_4);$

\bigskip

2.) Bifundamentals of $(2_i, 4_j)$ type and their complex conjugates

$(2_1, 4_2), (2_1, 4_3), (2_3, 4_1), (2_3, 4_3), (2_5, 4_1), (2_5, 4_2);
(2_2, 4_2), (2_2, 4_3), (2_4, 4_1), (2_4, 4_3), (2_6, 4_1), (2_6, 4_2); $
$ (2_7, 4_2), (2_7, 4_3), (2_9, 4_1), (2_9, 4_3), (2_{11}, 4_1), (2_{11},
4_2); (2_8, 4_2), (2_8, 4_3), (2_{10}, 4_1), (2_{10}, 4_3), (2_{12}, 4_1),
(2_{12}, 4_2)$ + complex conjugates.

\bigskip

3.) Bifundamentals of type $(4_i, \bar{4}_j)$ with $i\neq j $ and
$i,j=1,2,3$.

It is now easy to see that the one loop $\beta$ functions vanish for
all $SU(2)_a$ and $SU(4)_i$ groups as desired. This is of course not a
sufficient condition; however, we will assume that conformality condition
is satisfied to all loops.

Now let us note that all the multiplets in a.), b.) are vectorlike and
become massive at the conformlity scale. The required generations must
therefore come from the rest of the multiplets. For this purpose, let us
first break the $SU(4)_3$ group completely and consider the low energy
$SU(4)_c$ group as the direct sum of $SU(4)_1\times SU(4)_2$; also
aniticipating the discussion of $sin^2\theta_W$ in the next section, let
us consider the $SU(2)_L$ group as the direct sum of $SU(2)_m$'s for $m=
1,2,7,5$ and the $SU(2)_R$ as the sum of the rest of the $SU(2)_m$'s.
Then, three generations surviving at low energies correspond to the
``primordial'' representations $(4, 2_1)\oplus (4, 2_2)\oplus (4, 2_7)$
are left $SU(2)_L$ doublets
and $(\bar{4}, 2_3)\oplus (\bar{4}, 2_4)\oplus (\bar{4}, 2_9)$ are right
$SU(2)_R$ doublets, where we
have denoted $4 = 4_1 \oplus 4_2$. The multiplets of $(4_3, 2_k)$ and
$(\bar{4}_3, 2_l)$ type become vector like aftercomplete breakdown of
$SU(4)_3$ and pick up mass of order 100 TeV. For instance, $(2_8, 4)$ and
$(2_{10}, \bar{4})$ pair up to become massive via the Higgs coupling
$(2_8, 4)(2_{10},\bar{4})(2_8, 2_{10})_{H}$. One can write similar
couplings involving the fermions of $SU(4)_3$ that make all of them
massive.

\section{Prediction for $sin^2\theta_W(M_Z)$}

Using the standard evolution equations for nonsupersymmetric standard
model\cite{unif} and assuming as just noted the relations between the
$SU(4)_c\times SU(2)_L\times SU(2)_R$ gauge couplings and the unified
``static'' coupling $\alpha_U$, for the SU(2)'s to be
\begin{eqnarray}
\alpha^{-1}_{2L}(M_U)=p\alpha^{-1}_U, \nonumber \\
\alpha^{-1}_{2R}(M_U)=q\alpha^{-1}_U, \\ 
\alpha^{-1}_{4c}(M_U)=2r\alpha^{-1}_U, \nonumber
\end{eqnarray}
we can relate the observed standard model couplings at the scale $M_Z$ to
$\alpha_U$ and the scale of unification $M_U$. For this purpose, let us
remind the reader\cite{unif} that in the single scale breaking model that
we are considering, the hypercharge gauge coupling $\alpha^{-1}_1$ 
(suitably normalized) is related at the unification scale to the $SU(2)_R$
and the $SU(4)_c$ couplings by the relation
\begin{eqnarray}
\alpha^{-1}_{1}=\frac{2}{5}\alpha^{-1}_{4c} + \frac{3}{5}\alpha^{-1}_{2R}.
\end{eqnarray}
Using Eq. (1) and (2) and the beta functions of the standard model, we
can express $sin^2\theta_W$ and the QCD fine structure constant
$\alpha_s(M_Z)$ in terms of $\alpha_U$ and the unification scale $M_U$
(expressed below as $y=\ln \left(\frac{M_U}{M_Z}\right)$) as follows:
\begin{eqnarray}
\sin^2\theta_W(M_Z)=
\frac{p - (19/12\pi)y\alpha_U}{p+q +\frac{4}{3}r +
(11/6\pi)y\alpha_U}; \\
\nonumber
\alpha^{-1}_s(M_Z)=2r\alpha^{-1}_U-\frac{7}{2\pi}y.
\end{eqnarray}
Using these formulae and using $\alpha_s(M_Z)=0.12$, we find that for
$M_U \simeq 100$~TeV we get $\sin^2\theta_W(M_Z) \simeq 0.227$ for $p = 4$ 
and the simplest case $r = 2$ in the one loop
approximation. This is close to the central value from present
experiments\cite{pdg}. Clearly Eq.(3) with $p=4$ would give $\sin^2\theta_W
= 3/11$
for the too-low conformality scale $M_U=M_Z$ 
but this value is beautifully corrected to agree more closely
with experiment by the renormalization-group running between $M_Z$ 
and the conformality scale $M_U \simeq 100$~TeV.
Since $p$ can take only discrete integer values, we
find this remarkable that the prediction for $\sin^2\theta_W$ is close
to the observed value. 

\bigskip
\bigskip

As noted below, 100 TeV is not far from the lowest
phenomenologically allowed scale for the $SU(4)_C$ breaking scale 
derived from present upper limits on the branching ratio for the process
$K_L\rightarrow \mu^+e^-$\cite{desh}. 

\section{Other phenomenological issues:}
Once one has quark lepton unification at a scale near a 100 TeV, there are
several phenomenological issues, one has to deal with: first is the
splitting between quarks and leptons; the second question is how to
understand the smallness of neutrino masses and finally the question of
rare processes. We address them one by one:

\vskip0.3in
\noindent{\it Quark-lepton mass splitting}:
\vskip0.3in

There are two issues that need to be addresses here: first is how the
masses of different families of fermions arise and second how the quark
masses are split from the lepton masses despite the SU(4) unification
group. We do not address the details of the first question. Our view is
that the chiral families arise as a linear combination of large number of
high scale fermion representations and in the process, the different
families could acquire different Yukawa couplings that may eventually
explain the family hierarchy. 

We now address the second question which is also a nontrivial one since
since the Higgs boson content of the model
is severely limited by the requirements of conformal invariance. The 
allowed Higgs multiplets in addition to those stated earlier
(i.e. $\chi_{L,R}$ and $\phi$) are $\Sigma (15,1,1)$
(this multiplet can arise from $(4_1, 4_2)$ multiplets
when the diagonal sum is taken). It is then clear
that one can have higher dimensional operators of the form
$\bar{\Psi}_L\phi\Sigma\Psi_R/M$ where $\Psi = \left(\begin{array}{cccc}
u_1 & u_2 & u_3 & \nu \\ d_1 & d_2 & d_3 & e\end{array}\right)$ represents
the fermion multiplet. This operator splits the quarks from leptons and
provides a way to understand the quark-lepton mass splitting.

\newpage

\vskip0.3in
\noindent{\it Neutrino masses}
\vskip0.3in

The second question that one has to address has to do with the origin of
small neutrino masses. The usual seesaw mechanism\cite{seesaw} requires a 
high scale of about $10^{12}$ GeV or more for reasonable values of Dirac
masses for neutrinos. In the model, since the highest scale is only a
100 TeV, one must seek an alternative way to understand, why neutrinos are
so light. One possibility is to use a generalized seesaw that requires the
existence of gauge singlet fermions\cite{valle}. Suppose there are three
singlet neutrinos denoted by $\nu_{s,i}$ (i=1,2,3). Then one can write a 
mass matrix of the form for the neutrinos in the basis: $(\nu_e,
\nu_{\mu}, \nu_{\tau}; N_e, N_{\mu}, N_{\tau}; \nu_{s1}, \nu_{s2},
\nu_{s3})$
\begin{eqnarray}
{\cal M}_{\nu} = \left(\begin{array}{ccc} 0 & M_D & 0 \\
M^T_D & M_N & M_s \\ 0 & M^T_s & \mu_s\end{array}\right)
\end{eqnarray}
where $M_{D,N,s}$ are $3\times 3$ matrices. For typical values of the
masses in $M_D$ we assume 100 MeV since they originate at the weak scale
and ought to be of order of the charged lepton masses. $M_N$ is the
Majorana mass of the right handed neutrinos and is expected to be order
100 TeV as are the elements of $M_s$. As far as $\mu_s$ is concerned,
since the theory breaks lepton number (B-L to be precise), radiative
corrections can in conjunction with higher dimensional terms can generate
small masses for $\nu_s$ if we assume that they also have lepton number
like their electroweak counterparts. The effective ``active'' neutrino
masses are then given by:
\begin{eqnarray}
{\cal M}_{\nu}\simeq M^T_DM^{-1}_N\mu_sM^{-1}_N M_D
\end{eqnarray}
One can estimate typical magnitudes of the entries in ${\cal M}_{\nu}$,
the active neutrino masses, to be of order $\sim $ eV to milli eV
depending on the magnitudes of entries in $\mu_s$. Clearly these values
are in the right range to be of interest for explaining the current
evidences for neutrino oscillations.

\vskip0.3in
\noindent{\it Rare decay constraints}
\vskip0.3in

A major phenomenological constraint on these models comes from
the presence of the gauge leptoquark bosons connecting quarks to leptons
which lead to rare decays of the K-meson to $\mu^+e^-$ channels. The
strength of
these couplings is given by $\sim g^2_4/M^2_U\sim 10^{-10}$ GeV$^{-2}$.
Present upper limits on $B(K_L\rightarrow \mu^+e^-)\leq 6\times 10^{-12}$,
imply $M_U\geq 400$ TeV, which is not far from the unification scale that
gives the correct order of magnitude for the $sin^2\theta_W$ as discussed
above. Besides our estimate of the unification scale is based on an one
loop calculation and could easily be uncertain by a numerical factor of 3
to  4.

\newpage

\section{Conclusion and discussions}

In this note, we have studied the phenomenological implications of a
new approach to grand
unification that uses presumed conformal invariance of a nonsupersymmetric
theory derived from type IIB strings comapctfied on an ADS/$S_5$. We find
it interesting that there is only one model based on that uses
compactification on $S_5/\Gamma$ where $\Gamma = Z_3\otimes D_4$ group
which has three families. The gauge group of this model is 
$SU(4)_C \times SU(2)_L\times SU(2)_R$. 
We calculate the $sin^2\theta_W$ for this model
in the one loop approximation and find it to be $0.227$ which is in very
good agreement with the observations. We choose the scale to be in the 100
TeV range which is not far from that required by the rare $K_L\rightarrow
\mu^-e^+$ decay limits. We also discuss some other related
phenomenological issues.

\bigskip
\bigskip
\bigskip
\bigskip

\section*{Acknowledgements}

The work of PHF and SS is supported by a grant from US Department of
Energy DE-FG02-97ER-41036 and the work of RNM is
supported by a grant from the National Science Foundation under grant
number PHY-9802551.

\end{document}